\documentstyle[prl,aps,epsfig,multicol]{revtex}
\begin{document}
\def\be{\begin{equation}}
\def\ee{\end{equation}}
\def\bearr{\begin{eqnarray}}
\def\eearr{\end{eqnarray}}
\def\tc{$T_c~$}
\def\tcl{$T_c^{1*}~$}
\def\c2{ CuO$_2~$}
\def\ruo{ RuO$_2~$}
\def\lsco{LSCO~}
\def\bi{bI-2201~}
\def\tl{Tl-2201~}
\def\hg{Hg-1201~}
\def\sro{$Sr_2 Ru O_4$~}
\def\rc{$RuSr_2Gd Cu_2 O_8$~}
\def\mgb{$MgB_2$~}
\def\pz{$p_z$~}
\def\ppi{$p\pi$~}
\def\sqo{$S(q,\omega)$~}
\def\tperp{$t_{\perp}$~}
\def\srb{$SrB_{6}$~}
\def\cab{$CaB_{6}$~}
\def\bab{$BaB_{6}$~}

\def\b6{$B_6$~}

\title{Theory of High \tc Ferromagnetism in \srb family:\\
A case of Doped Spin-1 Mott insulator in a Valence Bond Solid Phase}

\author{ G. Baskaran \\
Institute of Mathematical Sciences\\
C.I.T. Campus,
Madras 600 113, India }

\maketitle
\begin{abstract}

Doped divalent hexaborides such as $Sr_{1-x}La_xB_6$ exhibit 
high \tc ferromagnetism. We isolate a degenerate pair of $2p$-orbitals 
of boron with two valence electrons, invoke electron correlation and 
Hund coupling, to suggest that the undoped state is better viewed 
as a spin-1 Mott insulator; it is predicted to be a type of 
3d Haldane gap phase with a spin gap $\sim 0.1~eV$, 
much smaller than the charge gap of $ > 1.0~eV$ seen in ARPES.
The experimentally seen high \tc `ferromagnetism' is argued to be 
a complex magnetic order in disguise - either a 
canted 6-sublattice AFM ($\approx 120^0$) order or its 
quantum melted version, a chiral spin liquid state, arising from
a type of double exchange mechanism.  

\end{abstract}

\begin{multicols}{2}[]

The observation\cite{young} of high \tc`ferromagnetism' in lightly 
doped \srb family is a great surprise in condensed matter physics 
in recent times; neither $Sr$ nor $B$  are known to participate 
in magnetism. Another surprise that followed was a 
high \tc superconductivity in \mgb, a diboride. Electron deficient 
$B$ is known to form molecules and solids with varying ligancy, from 
the stable doubly charged octahedral $(B_6H_6)^{2-}$ molecule to 
a metallic phase of boron, that has $B_{12}$ icosahedral cluster as 
a basic building block. How a 2p atom like boron in solid state manages 
to achieve high \tc superconductivity as well as high \tc ferromagnetism 
is a fascinating question.

There is sustained effort to understand ferromagnetism in the 
\srb family. Some of the new experimental results 
\cite{hbr.exp1,hbr.exp2,hbr.exp3,hbr.exp4,hbr.terashima,hbr.exp5} 
are intriguing and can not be explained in a satisfactory fashion
by the existing 
theories\cite{hbr.ceperley,hbr.russians,hbr.rice,hbr.varma,hbr.gorkov,hbr.ichin,hbr.pointdefect}
including the popular excitonic 
instability\cite{hbr.russians,hbr.rice,hbr.varma,hbr.gorkov,hbr.ichin}
scenario.
Notable among them are the recent ARPES and x-ray emission 
studies\cite{hbr.exp5}, that show a gap $> 1~eV$ at the Fermi level for
pure \srb. A gap of $\sim 0.8~eV$ is also predicted by an electronic 
structure calculation\cite{hbr.gw} using GW method that attempts 
to incorporate correlation effects in earlier 
approaches\cite{hbr.longuit,hbr.hasegawa,hbr.sandro}. 

In the present letter we propose that it is advantageous and perhaps
also correct to view the \srb family as a spin-1 Mott insulator, in view
of strong coulomb repulsions and Hund coupling among two electrons in two
degenerate valence orbitals of boron. However, at the outset we should
point out that the difference between a band and Mott insulator,
in an {\em even electron number} (per unit cell) insulator (with no
magnetic order)
like ours, is {\em quantitative} in the sense it is largely determined
by whether $E_T << E_c$ or not. Here $E_T$ is the spin gap or the 
lowest magnetic triplet exciton energy and $E_c$ is the charge
gap(as seen by ARPES for example).  

Our present proposal is consistent with all the known experimental
results and further certain novel predictions, which 
can be experimentally tested, follow in a natural fashion.
i) \srb is a spin-1 Mott insulator, with two electrons in a doubly 
degenerate $2p$-orbitals providing a spin-1 moment. ii) 
Antiferromagnetic coupling among the localized spin-1 moments, arising 
from kinetic exchange leads to a type of {\em 3d  Haldane 
gap or Majumdar-Ghosh phase with unbroken lattice symmetry}(figure 1).  
The singlet valence bonds are `frozen' at the $B-B$ bond 
bridging two neighboring \b6 octahedra. This phase has a small 
spin gap $E_T \sim 0.1~eV$ (this is not incompatible 
with a diamagnetic behavior of the insulating \srb). 
iii) Doping liberates, through a form of double exchange mechanism, 
a canted 6-sublattice 
AFM ($\approx 120^0$) order or a chiral spin liquid 
state,  $\langle{\bf S}_{i\alpha} \cdot ({\bf S}_{i\beta} \times 
{\bf S}_{i\gamma})\rangle \neq 0$ both with a small ferromagnetic 
moment.  Our predictions could be tested by neutron scattering 
or other low energy probes. Two of our robust predictions are a 
small spin gap, $E_T \sim 0.1~eV$ and at the least a well 
developed $120^0$ spin correlations (figure 1) inside a \b6 octahedron.

The following two experimental facts about \srb, when taken 
together are striking and gives an important clue for our model 
building. i) As revealed by recent ARPES 
experiments\cite{hbr.exp5} the parent insulator has a 
charge gap $ >1~eV$ ii) The insulating 
paramagnetic ground state is very fragile (unlike a band insulator)  
and a small doping in $Sr_{0.995}La_{0.005}B_6$, that adds half a percent 
of carrier per $La$ atom, produces a ferromagnetic phase (with
a small moment $\approx 0.07 \mu_{B}$ per La atom) with a large 
\tc $\approx 600 - 900 K$. 

Some key quantum chemical information about the \srb family
that we will use in building our theory are: i) each $B$ atom has 
two valence electrons and two degenerate valence $p$-orbitals,
 ii) the intra atomic Hund coupling  is 
$\approx 1.5 eV$, iii) the unscreened 
Hubbard U for the $2p$ orbitals of $B$ can be as large 
as $ 8 - 10~eV$.
iv) The nearest neighbor inter octahedral $B-B$ distance is smaller 
than that inside an octahedron by about 5 percent. 

We first briefly present the conventional electronic 
structure\cite{hbr.longuit} description 
of $B_6$ cluster in \srb. $B_6$ octahedra are covalently
bonded to form a simple cubic lattice. There is nearly complete charge
transfer from Sr to $B_6$ cluster:$Sr^{2+} {B_6}^{2-}$. Ignoring the 
core orbitals of $B$ as well as $Sr^{2+}$ we are left with one 2s
and three 2p orbitals per $B$ atom. There are 20 electrons in these
24 $B$  orbitals in every octahedron. The 24 orbitals can be separated 
into two sets:
i) $sp^1$ hybrids (two per $B$ atom) that are along the body diagonal
of the $B_6$
octahedra and ii) two $p$-orbitals per $B$ atom that are tangential
to the octahedra (inset in figure 1).  
An s-like combination of the six $sp^1$ 
hybrids that are pointing at the octahedral center form a very
strong six center bond with a binding energy of about 15 eV. This bonding
state takes two electrons and is primarily responsible for the stability
of the octahedra. The $sp^1$ orbital pointing radially outwards 
strongly hybridize with the corresponding orbital of neighboring 
octahedra and 
results in a stable covalently bonded cubic network of $B_6$ 
octahedra. These bonding states take away 6 electrons per $B_6$
octahedra.  {\em We are now left with 12 electrons and 12 $p$-orbitals 
per octahedron, i.e.,two tangential $2p$ orbitals and two electrons
per B atom}.

In the first band structure calculation\cite{hbr.longuit}
 for \srb family, Longuet-Higgins 
and Roberts, using the 12 $p$-orbitals, form a set of four
triply degenerate molecular orbitals -  
$t_{1u}, t_{2g}, t_{1g}$ and $ t_{2g}$. 
(For simplicity we will ignore the hybridization of $t_{1u}$ with symmetry 
adapted $sp^1$ orbitals that leads to $t'_{1u}$ and $t"_{1u}$). 
These \b6 cluster orbitals overlap to produce 6 bonding and 
6 anti bonding Bloch bands. The 6 bonding bands are completely filled
by 12 electrons to produce a band insulator.  Later 
workers\cite{hbr.hasegawa,hbr.sandro}
emphasized the mixing of anion d-states and showed that within 
band theory it reduces the band gap to nearly zero value at the X-points
in k-space. 

In the absence of any electron-electron interaction the energy 
difference between the top most $t_{1g}$ and and bottom most 
$t_{1u}$ molecular orbitals is $\sim 10~eV$. 
This is a measure of the kinetic energy of 
delocalization per electron in the slater determinant state within 
a \b6 cluster. This energy is comparable to the Hubbard U $\sim 8~-
10~eV$.  In addition we have a ferromagnetic Hund coupling between 
two $p$-orbitals of a $B$ atom, $J_H \sim 1.5~eV$. As the kinetic
energy of delocalization within the cluster is comparable to the energy 
increase from Hubbard U and Hund coupling $J_H$, the low lying cluster 
eigen states are strongly perturbed by many body effects, to 
the extent the simple \b6 molecular orbitals and their slater 
determinant states mostly loose their relevance. 

{\em Thus a natural starting point to understand the low energy 
physics of the cluster is the Mott localization of two electrons 
and form a spin-1 boron moment}. Our approach is
similar to restricting oneself to the valence band basis in the case
of carbon $p\pi$ bonded system such as benzene (with very similar quantum
chemical parameters), where it is known to work very well; 
ours is a generalization of valence bond basis approach
to the case of two $p$-orbitals per site. Further the 
inter \b6 cluster hoping does not modify this localized picture 
significantly, as the inter molecular orbital hopping matrix element 
between neighboring clusters is small $\sim 0.25~eV$ (that leads to 
a band width of $ 12 \times 0.25 \approx 3~eV$, as seen in band 
structure results).   

We have formalized the above by starting from a doubly degenerate 
3 dimensional Hubbard model containing an average of one electron
per orbital:
\bearr
H  =  -\sum_{\langle ij;\alpha\beta;\mu\nu \rangle } 
t_{ij}^{\alpha\beta;\mu\nu} 
& C^{\dagger}_{i\alpha\mu\sigma} & C^{}_{j\beta\nu\sigma}+
U\sum_{i\alpha\mu}n^{}_{i\alpha\mu\uparrow}n^{}_{i\alpha\mu\downarrow}
\nonumber \\
~&-&~J_H\sum_{i \alpha} {\bf s}_{i \alpha 1}\cdot{\bf s}_{i \alpha 2}
\eearr
Here $t_{ij}^{\alpha\beta;\mu\nu}$ represents the nearest neighbor
hopping integrals; i denotes the centers of octahedra that form a 
simple cubic lattice, $\alpha = 1,..6$ denotes a $B$ site within 
an octahedron and $\mu = 1,2$ denotes the two degenerate $p$-orbitals.  
The operator ${\bf s}_{i \alpha \mu}$ is a spin half operator
of an electron in the $\mu$th $p$-orbital of $\alpha$'th $B$ site 
in the i-th octahedron.

Our Hubbard model contains three types of nearest neighbor hopping 
matrix elements $t_{\pi1}$, $t_{\pi2}$, and $t_{\pi\sigma}$
shown in the inset in figure 1; their values range over 
$1~eV$ to $1.5~eV$.  As mentioned earlier U $\sim 8~eV$  
and $J_H \sim 1.5~eV$.

To understand the low energy spin dynamics we perform a kinetic
exchange perturbation theory and get a 3 dimensional spin-1 
Hamiltonian for our Mott insulator:
\be
H_s = J_1\sum_{i,\langle \alpha \beta \rangle} {\bf S}_{i,\alpha}\cdot
{\bf S}_{i, \beta} +  J_2\sum_{{\langle ij ; \alpha \beta \rangle}}
{\bf S}_{i,\alpha}\cdot {\bf S}_{j,\beta}
\ee
The first term represents the \b6 cluster spin Hamiltonian and second 
the octahedral bridge spin coupling.  Here ${\bf S}_{i\alpha}$ is the 
spin-1 operator of the $\alpha$'th $B$ atom in the i-th \b6 
octahedron. $J_1$ and $J_2$ are intra cluster and inter cluster kinetic 
exchange integrals. A direct kinetic exchange perturbation
theory gives a large antiferromagnetic coupling and it needs to be 
corrected by subtracting a direct or `potential exchange' (arising
from non-orthogonality of the nearest neighbor orbitals) that
favors ferromagnetic alignment. We have estimated the potential 
exchange by looking at some experimental results and quantum chemical 
calculations for the related carbon systems. After making this 
subtraction we estimate $J_1, J_2 \sim 0.1$ to $ 0.3~eV$ and
$J_1 < J_2$.
In view of the approximate nature of our estimates, we will keep
$J_1$ and $J_2$ as parameters to be determined experimentally.

Now we derive the phase diagram for our spin Hamiltonian 
as a function of its only parameter, the dimensionless ratio 
$\frac{J_2}{J_1}$ (figure 2). 
It is convenient to rewrite the Hamiltonian (equation 2) as 
\bearr
H_s & = & \frac{J_1}{4}\sum_{i,\langle \alpha \beta \gamma \rangle} 
({\bf S}_{i,\alpha} + {\bf S}_{i, \beta} + {\bf S}_{i, \gamma})^2
\nonumber \\ 
& + &  \frac{J_2}{2} \sum_{{\langle ij ; \alpha\beta \rangle}}
({\bf S}_{i,\alpha} + {\bf S}_{j,\beta})^2 - N(12J_1 + 6J_2) ,
\eearr
\begin{figure}[h]
\epsfxsize 8cm
\centerline {\epsfbox{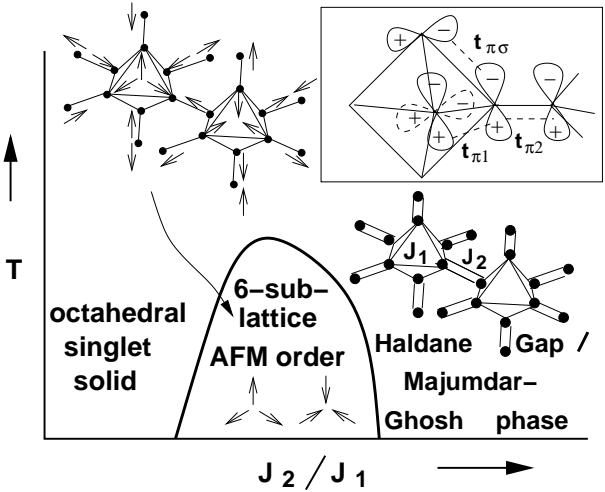}}
\caption{Schematic phase diagram of our Boron spin-1 Hamiltonian.
Doubles lines denote `frozen' valence bonds at the octahedral 
bridges. Inset shows the `tangential' boron $2p$ valence orbitals 
and 3 types of hopping matrix elements.} 
\end{figure}
where N is the total number of octahedra. 
The first term of equation (3) represents the sum over triangles 
of three spins   
forming the eight faces of a \b6 octahedron. Each term of 
equation 3  
is a positive operator with eigen values $\geq 0$. It follows 
immediately that the classical ground state of the above Hamiltonian
is a $120^0$ six-sublattice antiferromagnet for any $J_1,J_2 > 0$. 
The spin pattern within an octahedron is shown in figure 2. 
Spins pairs at opposite corners of an octahedron are 
parallel. Three such diagonal pairs within an octahedron  are coplanar 
and at an angle of $120^0$, among themselves. This defines a spin  
chirality (vorticity) $\pm 1$ for an octahedron. 
The cubic lattice of octahedra form two cubic sublattices with 
corresponding spins exactly anti-parallel, making it a 6-sublattice 
planar antiferromagnet (figure 1).  The net result is that the bridge spin 
coupling is not frustrated, but {\em the twelve 
nearest neighbor coupling within an octahedron are frustrated}.
In spite of the spin reversals the chirality has the same sign in
all the octahedra; i.e., we have a ferromagnetic chiral order.
Thus our ground state exhibits a discrete two fold chirality (Ising like) 
degeneracy in addition to the global SU(2) spin rotational degeneracy.   

For $ \frac{J_2}{J_1} \approx 1$ the classical ground state 
exhibited above is stable and survives spin wave fluctuations.
However, when $ \frac{J_2}{J_1} >> 1$ or $ \frac{J_2}{J_1} << 1$ 
quantum fluctuations destabilize the classical ground state and 
we get singlet ground states with a finite spin gap.

When $ \frac{J_2}{J_1} << 1$ each \b6 cluster has an unique  
singlet ground state. There is a finite gap $\sim J_1$ for spin-1 
excitation.  The couplings $J_2$  between neighboring octahedra 
through the bridge spins reduce this gap by virtual excitations. 
We have estimated by perturbation theory that the gap survives until  
$ \frac{J_2}{J_1} \sim {1\over 2} $. We call this phase as the 
octahedral singlet solid (OSS). 

When $ \frac{J_2}{J_1} >> 1 $ the AFM order is destabilized;
the octahedral bridge pairs become non-degenerate singlets 
with a gap of $\sim J_2$ for spin-1 excitations. Spin couplings
within \b6 cluster reduce this gap by virtual excitations 
and the gap vanishes when we decrease $\frac{J_2}{J_1}$ to a value 
$\sim 2$. This {\em valence bond solid} phase is a 3 dimensional 
realization of spin-1 Majumdar-Ghosh phase, that also retains 
the lattice symmetry. Further, this phase is also continuously 
connected to an AKLT\cite{aklt} type of 3d Haldane gap phase, 
with quantum fluctuating effective spin-3 moments at \b6 
clusters\cite{bas.aklt}

Where is \srb family in the above phase diagram ? It is 
unlikely that undoped \srb family has long range AFM order.
The shortness of the octahedral bridge $B-B$ distance by about
5 percent, compared to nearest neighbor $B-B$ distance within an 
octahedra seen in all the 
three hexaborides, \srb, \cab, \bab could also arise, apart from 
some quantum chemical reasons, from the formation of a `frozen' 
singlet (valence bond) between the $B$ spin-1 moments of the
octahedral bridge pair. So it is very likely that the strong 
quantum fluctuations of the spin-1 system and the fact that
$\frac{J_2}{J_1} \geq 1$ 
in the \srb family is keeping it in a valence bond solid phase.

Now we discuss how very small doping leads to a ferromagnetic order 
with a large curie temperature. The analysis of the doped situation 
starting from the Hubbard model (equation 3) is hard compared to 
the undoped case.  However, our preliminary analysis indicates 
very rich possibilities, all arising from the Mott insulator 
parentage and the `vicinity' to an antiferromagnetic order. 

We believe that the observed small moment gives us an important
clue as to the origin of magnetism. {\em A natural way in which a small 
moment can arise is by canting of an existing antiferromagnetic order}.
Our valence bond solid phase, being close to the 6-sublattice 
AFM phase in the phase diagram, should have short range 
antiferromagnetic correlations. Further if doped carrier 
delocalization decreases the effective $\frac{J_2}{J_1}$ of our
boron spin Hamiltonian, we my be pushed into the 6-sublattice
AFM phase: we argue below that this is likely to happen. 
\begin{figure}[h]
\epsfxsize 9cm
\centerline {\epsfbox{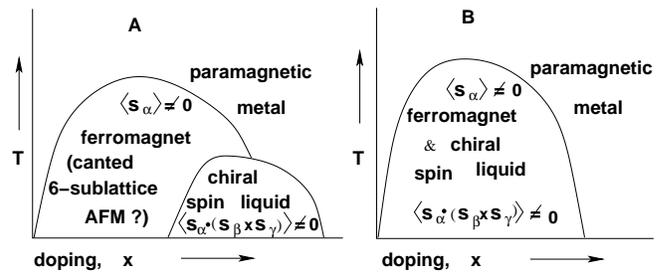}}
\caption{Two Possible Schematic Phase diagrams in the $x$-$T$ plane. 
The experimentally observed `ferromagnetic' phase is suggested to be 
a canted 6-sublattice antiferromagnetic or a chiral spin liquid
phase} 
\end{figure}
In a double exchange\cite{hbr.pwahasegawa}
 process, an added electron prefers higher 
total spins such as 2 or 1 between neighboring $B$ spin-1 pairs, 
in order to satisfy Hund's rule during delocalization; {\em singlet pairs 
are bottle necks for doped carrier delocalization}. In the VBS phase
such singlet amplitudes
among neighboring spin pairs within an octahedron are relatively low: 
total spin values $2,1$ and $0$ of a pair occur with probabilities 
$\sim \frac{5}{9}$,$\frac{3}{9}$ and  $\frac{1}{9}$; on the other hand
the singlet amplitude is larger $\sim 1$  at an octahedral bridge pair.
Carrier delocalization thus will selectively project out large local 
singlets from the bridge pairs; i.e., it effectively adds a local term, 
a singlet projector, 
$ x t_{\pi2} ({\bf S}_{i\alpha} + {\bf S}_{j\beta})^2$,
to the singlet dominated octahedral bridge pair. Thus $J_1$ and $J_2$
of our spin Hamiltonian (equation 2) get modified in a {\em state
dependent fashion} to effective $\tilde{J}_2(x) \approx J_2 - x t_{\pi2}$ 
and $\tilde{J}_1(x) \approx J_1$.

The above results in a linear decrease in the ratio with doping $x$: 
$\frac{\tilde{J}_2(x)}{\tilde{J}_1(x)} \sim \frac{J_2}{J_1} 
- x \frac{t_{\pi2}}{J_1}$. Thus doping may allow us to enter(in
figure 1) the 6-sublattice AFM phase (with a small canting induced 
ferromagnetism, to be discussed below) or a quantum melted 
version with long range chiral order. The scale of maximum \tc 
in our picture is roughly the maximum scale of \tc of the insulating 
part of the phase diagram, determined by $J_1$ $J_2$: 
$k_BT_c \approx J_1,~J_2$, which can be easily as large as 900 K.

Once we establish a well developed short range or long range 
AFM order canting is easily obtained in the double exchange 
mechanism, as explained by de Gennes in 1960 in the context of
doped manganites. This is explained by minimizing the sum of 
the exchange energy of an octahedral bridge pair spins, for
example, and the 
Anderson-Hasegawa double exchange term\cite{hbr.pwahasegawa}:
\be
  E \approx J_2 \cos \theta - t_{\pi2}~ x \cos \frac{\theta}{2}
\ee
Here $\theta$ is the angle between the two spins.
The energy minimum occurs at $\cos \frac{\theta}{2} = 
\frac{t_{\pi2}~x}{4J}$, i.e., $\theta \approx 180^0  
- \frac{t_{\pi2}~x}{4J}$,
rather than at $\theta = 180^0$. This leads to a small moment
of $ \approx 0.9~\mu_B$ per formula unit, close to what is seen 
experimentally. 

The preexisting planar chirality (vorticity) in our 6-sublattice 
AFM order, elaborated earlier, gives us a novel possibility 
of a finite non-planar chirality order\cite{kalmayer}  
$\langle{\bf S}_{i\alpha} \cdot ({\bf S}_{i\beta} \times 
{\bf S}_{i\gamma})\rangle \neq 0$, through quantum fluctuations
arising from dopant carrier delocalization.
How canting, a net magnetic moment and  a non-planar chirality 
arise through quantum fluctuations is nicely 
illustrated in the following example. Consider 
three spin-half moments coupled antiferromagnetically:
$H = J  (
{\bf S}_{1} \cdot {\bf S}_{2} + {\bf S}_{2} \cdot {\bf S}_{3} + 
{\bf S}_{3} \cdot {\bf S}_{1} )$ has a 
$120^0$ classical planar ground state ($S_z = 0$)  
with a two fold planar chirality 
degeneracy. However, the exact ground states have non-planar 
chirality $\langle{\bf S}_{1} \cdot ({\bf S}_{2} 
\times {\bf S}_{3})\rangle_{G} = \pm 2\sqrt3$ and a net 
spin =$\frac{1}{2}$ moment, both arising from quantum fluctuation
induced canting. One should not also rule out the possibility of
nematic order such as $\langle{\bf S}_{\alpha} \times 
{\bf S}_{\beta}\rangle_{G} \neq 0 $ in view of the complexity
in our system. 

Thus canting, weak moment ferromagnetism and development of 
non-zero chiral order parameter 
$\langle{\bf S}_{i\alpha} \cdot ({\bf S}_{i\beta} \times 
{\bf S}_{i\gamma})\rangle \neq 0$ are all tied together.
As the energy scales and the origin of the chirality 
stiffness and the 6-sublattice AFM stiffness are different   
the career delocalization can also quantum melt the long
range AFM order but leave the chirality order intact.
In this case we will have a chiral spin liquid with a 
weak ferromagnetic moment. Figure 2 gives two schematic phase 
diagrams in the $x-T$ plane.

Important questions as to why ferromagnetism occur in a narrow range 
of doping $x$, the complex transport, and sharpening our various 
estimates and heuristic arguments by many body methods remains to be done. 

I thank P.W. Anderson for an encouraging discussion.

\end{multicols}
\end{document}